\documentclass[iop,apj,useAMS,usenatbib]{emulateapj}
\usepackage{multirow}
\usepackage{multirow}
\usepackage{enumerate}
\usepackage{amsmath,epsfig,url,natbib}
\usepackage{graphicx,epstopdf,float,color}
\usepackage{courier}
\usepackage[hang,flushmargin]{footmisc}
\usepackage{hyperref}
\usepackage{gensymb} % for degree symbol
\usepackage[flushleft]{threeparttable}
\usepackage[thinlines]{easytable}
\usepackage{longtable,natbib}
\hypersetup{colorlinks=true, citecolor=blue, filecolor=magenta, urlcolor=cyan, linkcolor=blue}
\def\lsim{\mathrel{\rlap{\lower4pt\hbox{\hskip1pt$\sim$}}
    \raise1pt\hbox{$<$}}}                % less than or approx. symbol
\def\gsim{\mathrel{\rlap{\lower4pt\hbox{\hskip1pt$\sim$}}
    \raise1pt\hbox{$>$}}}                % greater than or approx. symbol

\submitted{}
\begin{document}

\title{{{{L}\lowercase{ens}{F}\lowercase{low}}: A Convolutional Neural Network in Search of Strong Gravitational Lenses}}

\author{Milad Pourrahmani}
\author{Hooshang Nayyeri}
\author{Asantha Cooray}

  \affil{Department of Physics and Astronomy, University of
  California Irvine, Irvine, CA}

\journalinfo{Submitted to the Astrophysical Journal}
\submitted{}
\begin{abstract}
We have entered the era of big data astronomy. Sky surveys such as the LSST, Euclid, and WFIRST will produce more imaging data than humans can ever analyze by eye. The challenges of designing such surveys are no longer merely instrumentational, but they also demand powerful data analysis and classification tools that can identify astronomical objects autonomously. To gradually prepare for the era of autonomous astronomy, we present our machine learning classification algorithm for identifying strong gravitational lenses from wide-area surveys using convolutional neural networks; LensFlow. We train and test the algorithm using a wide variety of strong gravitational lens configurations from simulations of lensing events. Images are processed through multiple convolutional layers which extract feature maps necessary to assign a lens probability to each image. LensFlow provides a ranking scheme for all sources which could be used to identify potential gravitational lens candidates by significantly reducing the number of images that have to be visually inspected. We further apply our algorithm to the \textit{HST}/ACS i-band observations of the COSMOS field and present our sample of identified lensing candidates. The developed machine learning algorithm is much more computationally efficient than classical lens identification algorithms and is ideal for discovering such events across wide areas from current and future surveys such as LSST and WFIRST.
\end{abstract}

\section{Introduction}\label{intro}

Gravitational lensing, a prediction of Einstein's general theory of relativity, is a very powerful tool in cosmological studies. It has been used extensively to understand various aspects of galaxy formation and evolution (e.g. \cite{refsdal1964gravitational, blandford1992cosmological, nayyeri2016candidate, postman2012cluster, atek2015new}). This involves accurate cosmological parameter estimation \citep{treu2010strong}, studies of dark matter distribution from weak gravitational lensing events \citep{kaiser1993mapping, velander2014cfhtlens}, black-hole physics \citep{peng2006probing} and searches for the most distant galaxies \citep{coe2012clash, oesch2015first}, among others. 

One of the main goals of observational cosmology is to constrain the main cosmological parameters that dictate the evolution of the Universe \citep{tegmark2004cosmological, komatsu2009five, weinberg2013observational}. Strong gravitational lensing has been utilized over the past few years to estimate and contain these cosmological parameters \citep{broadhurst2005strong, suyu2013two, suyu2014cosmology, goobar2016, angnello2017, more2017}. This is achieved through accurate lens modeling of such events and comparing the model predictions with observations (such as with observations of lensing induced time delays \citep{eigenbrod2005, treu2010strong, suyu2014cosmology,  rodney2016, treu2016}. In a recent study, for example, \citet{suyu2013two} used combined WMAP, Keck and {\it HST} data on gravitational time delays in two lensed sources to constrain the Hubble constant within $4\%$ in a $\rm \Lambda CDM$ cosmological framework.

One of the key aspects of gravitational lensing is its use as natural telescopes through boosting the observed signal and increasing the spatial resolution \citep{treu2010strong}. This is quite advantageous in searches for distant and/or faint objects at moderate observing costs and has been utilized extensively in various surveys in searches for such objects, the identification of which would not have been possible without it \citep{bolton2006sloan, heymans2012cfhtlens}. Given that the number of identified lenses for different classes of galaxies rises sufficiently due to better lens finding algorithms, by modeling the lenses and using spectra stacking techniques, we can better understand the physical and chemical composition of farther and fainter galaxies which in turn would excel our understanding of galaxy evolution \citep{wilson2017, timmons2016}. In the past few years, deep diffraction limited observations have also taken advantage of gravitational lensing to extend the faint end of the luminosity function of galaxies by a few orders of magnitude \citep{atek2015new} to produce the deepest images of the sky ever taken across multiple bands. Strong gravitational lensing events have been observed extensively in such surveys as galaxy-galaxy lensing in field surveys such as the Cosmic Assembly Near-infrared Deep Extragalactic Legacy Survey (CANDELS) \citep{grogin2011candels, koekemoer2011candels} and the Cosmological Evolution Survey (COSMOS) \citep{scoville2007cosmos, capak2007} or as cluster lensing from observations of nearby massive clusters \citep{postman2012cluster, treu2015grism, lotz2017frontier} with {\it Hubble} Space Telescope. These yield identification of the first generations of galaxies at $z>3$ (and out to $z\sim11$; \citealp{oesch2015first}) to study galaxy formation and evolution at the epoch of re-ionization. This was, in fact, one of the main motivations behind Hubble cluster lensing studies such as CLASH and Frontier Fields \citep{postman2012cluster, lotz2017frontier}. These magnification provided by the strong lensing could potentially raise the observed flux by as much as a factor of hundred (depending on configuration). The power of gravitational lensing could also be used in the detection of low surface brightness emission from extended objects such as far-infrared and radio emissions from dust and molecular gas at $z\sim2-3$. This is indeed one of the main techniques in observing these systems even in the era of powerful mm/radio telescopes such as ALMA. 

Strongly lensed galaxies are normally targeted and identified from dedicated surveys \citep{bolton2006sloan}. Traditionally these lens identifications are either catalog-based, in which lensing events are identified by looking for objects in a lensing configuration, or pixel based, with the search starting from a set of pixels. These lensing searches are normally computationally challenging in that individual pixels are constantly compared with adjacent ones and they could be biased towards a given population and/or brighter objects. Recent far-infrared wide area observations (such as with \textit{Herschel}) significantly advanced these searches for lensed galaxies by adopting a simple efficient selection technique of lensed candidates through observations of excessive flux in the far infrared (as an indication of strong lensing events supported by number count distributions; \citealp{nayyeri2016candidate, wardlow2012hermes}). However such surveys are also biased towards populations of red dusty star-forming galaxies (missing any blue lenses) and are not always available across the full sky (the \textit{Herschel} surveys that were targeted had $\rm \sim0.2-0.4\,deg^{-2}$ lensing events, much lower than expected from optical surveys). Given that tests of cosmological models require simple unbiased selection function, it is important to have a complete unbiased catalog of lensing events. 

\begin{figure} 
    \includegraphics[width=0.5\textwidth]{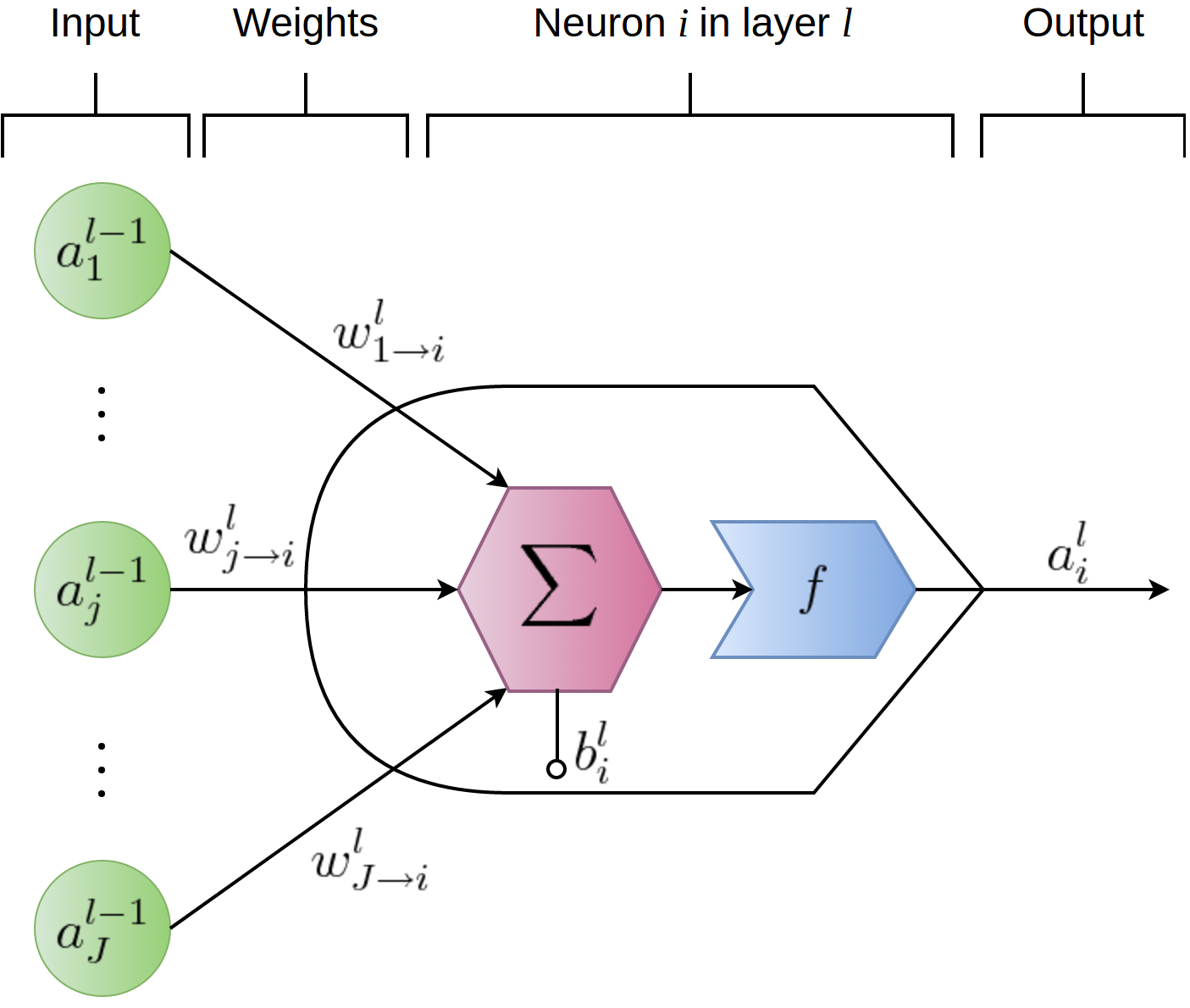}
    \centering
  \caption{Schematic representation of an artificial neuron. The weighted sum of the neurons in the previous layer (green circles), plus the internal bias of the neuron, are mapped as the output of the neuron by an activation function. This model is captured by Equation \ref{eq:activation}. During the learning process, weights and biases of the neurons will be adjusted to achieve the desired network output.}
\label{fig:neuron}
\end{figure}

We have entered the era of big data astronomy. Sky surveys such as the LSST, Euclid, and WFIRST will produce more imaging data than humans can ever analyze by eye. The challenges of designing such surveys are no longer merely instrumentational, but they also demand powerful data analysis and classification tools that can identify astronomical objects autonomously. Fortunately, computer vision has drastically improved in the last couple of years to make autonomous astronomy possible. The past couple of years has been the most exciting era in the field of machine learning (ML). Researchers from both the public and the private sectors have achieved landmarks in developing image recognition/classification techniques. One of the most exciting recent events in the ML community was the release of {\sc TensorFlow} by Google, a parallel processing platform designed for development of fast deep learning algorithms \citep{abadi2016tensorflow}. Packages like {\sc TensorFlow}, {\sc Caffe}, and others have enabled researchers to develop very complex and fast classification algorithms. Among these deep learning programs, ConvNets have deservingly received a lot of attention in many fields of science and industry in the past few years \citep{krizhevsky2012imagenet}. Complex ConvNets such as {\sc GoogleNet} and {\sc AlexNet}, which are publicly available, have achieved superhuman performance on the task of image classification. Google's {\sc TensorFlow} has made it possible to easily develop parallelized deep learning algorithms which if integrated with Google's Tensor Processing Units (TPUs), could address the data mining challenges in the field of astronomy. The field of astronomy and observational astrophysics should take advantage of these new image classification algorithms. For instance, our team has been working on developing LensFlow, a ConvNet that can be used to search for strong gravitational lenses. Our work will be publicly available on Github after publication.
 
Non-machine learning computer algorithms have been previously used for finding gravitational lenses \citep{alard2006, arc_finder2012}. For instance, \citet{arc_finder2012} use two algorithms called {\sc RingFinder} and {\sc ArcFinder}. The former uses color information and the latter detects arc-like pixels.  Initially, {\sc ArcFinder} polishes the images by convolving a smoothing kernel. For each pixel, an estimator of elongation is calculated by taking the ratio between the sum of the flux of a few pixels along the horizontal line and the maximum value of a few nearby pixels along the vertical line which pass through the pixel in hand. This process is repeated for all pixels and those with smaller than an specified elongation threshold are set to zero to create a sharp arc map. An arc map that satisfies thresholds on the arc properties such as the size and surface brightness will be selected as an arc candidate for further visual inspection. Unlike deep learning, such classical algorithms are not easily parallelized and they can be computationally more expensive depending on the deepness of the ConvNets used. They also require threshold tuning which may cause insensitivity to smaller arcs. However, they do not require a massive training dataset. Even though more challenging, creating a large dataset could eliminate biases toward certain arc-lens morphologies. As we will discuss in Section \ref{cosmos}, these algorithms do suffer from the same lens contaminants as ours. The hope is that machine has the capacity to improve its performance with better training datasets and improved architecture while remaining computationally efficient but such improvements are not trivial regarding classical algorithms.
 
Other researchers \citep{1st,2nd,3rd} also find deep learning a suitable solution for finding gravitational lenses. \citet{3rd} use residual ConvNets with 46 layers. Residual ConvNets are modified ConvNet that do not suffer from layer saturation as ordinary ConvNets do. After adding more than 50 layers, the accuracy of ordinary ConvNets no longer improves and the training becomes more challenging. \citet{he2016deep} were able to overcome this issue by providing residual maps in between layers, which has been employed by \citet{3rd}. They have simulated LSST mock observations in a single band and have trained and tested their network on these images. \citet{2nd}  have trained their ConvNet using multiple color bands and have applied it to Canada-France-Hawaii Telescope Legacy Survey. \citet{1st} have searched for lenses in Kilo Degree Survey by training their ConvNet on cataloged luminous red galaxies.

In our independently developed work, we focus on the morphology of the lenses and only rely on one color band, similar to \citet{1st} and \citet{3rd}. Our lens simulation method is very similar to \citet{1st} where we both merge simulated arcs with real images of galaxies to preserve the complexity of the physical data. In contrast to others, we do not discriminate against different sources found in the COSMOS field. Artifacts, starts, and other sources have been included in our training dataset so LensFlow can be directly applied to fields without a need for a catalog with galaxy type information. The deepness of our ConvNet is comparable to \citet{1st} and \citet{2nd} but it is shallower than \citet{3rd}. As mentioned in \citet{2nd}, the morphology of lenses are much simpler than the morphology of daily objects and human faces which extremely deep ConvNets are developed for. However, the cost to performance ratio of ConvNets with varying deepness has not been studied yet. The effectiveness of deeper ConvNets cannot be compared between ours (and \citealp{1st} ) and \citet{3rd} since they have not applied their algorithm to physical data. However, they have studied the change in the performance of their ConvNet by varying the Einstein radii and signal-to-noise ratio of their lenses.

This paper is organized as follow. In Section \ref{intro to ML}, we will explain the principal concepts underlying neural networks, supervised learning, and ConvNets. A supervised learning algorithm requires a large sample of labeled images, known as the training dataset.  In Section \ref{datasets}, we will discuss the procedure we have taken to create our training and testing datasets. Before feeding the images to a ConvNet, they must be normalized and should be enhanced. The details of these methods are discussed in Section \ref{normalization}. Section \ref{convnet arch} will lay out the architecture of LensFlow and Section \ref{acc} will illustrate  LensFlow's performance on the testing dataset. We will conclude by sharing and analyzing the scan results of the COSMOS field in Section \ref{cosmos}. Throughout this paper, we assume a standard cosmology with
$H_0=70\,\text{kms}^{-1}\text{Mpc}^{-1}$, $\Omega_m=0.3$ and $\Omega_\Lambda=0.7$. Magnitudes are in the AB system where
$\text{m}_{\rm AB}=23.9-2.5\times\text{log}(f_{\nu}/1\mu \text{Jy})$ \citep{oke1983secondary}. 

\begin{figure} 
    \includegraphics[trim=2cm 0cm 0cm 0cm, width=0.5\textwidth]{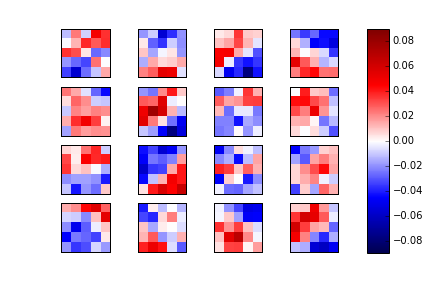}
    \centering
  \caption{Different filters used in the first convolutional layer. The pixels in each box represent the weights of a convolving neuron which are connected to a $5 \times 5$ region input image. As these filters convolve over the entire input image, they generate 16 feature maps. Red pixels have a positive contribution and blue pixels have a negative contribution toward the activation of the convolving neuron. These filters are helpful for edge and texture recognition.}
\label{fig:filters1}
\end{figure}

\begin{figure} 
    \includegraphics[width=0.5\textwidth]{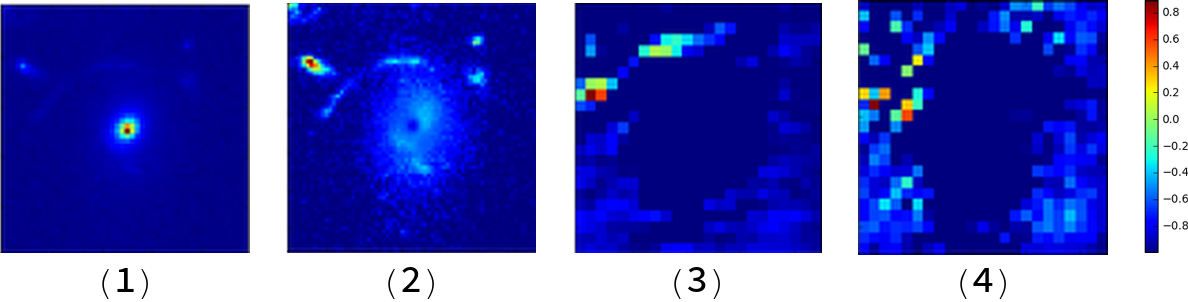}
  \caption{Two examples of convolutional layer feature maps. Image of a normalized physical lens has been shown in (1). After the normalizing and enhancing it with an anti-Gaussian function, the arc stands out (2). Feature map (3) shows an enhancement of the arc while feature map (4) highlights the edges along the anti-diagonal orientation. Such feature maps are to be further compared and analyzed in the fully-connected layers to determine if the image is lensed or not.}
    \centering
    \label{fig:feature_maps}
\end{figure}
 
 \section{Deep Learning Algorithms}\label{intro to ML}
Artificial neural networks are inspired by biological neurons. Just like biological neurons, artificial neurons receive input signals and send out an output signal to other neurons (see Figure \ref{fig:neuron}). The synaptic connections between neurons are known as weights and the output of a neuron is know as its activation. To reduce the computational time and simplify neural network models, neurons are placed in consecutive layers rather than having a connection with every other neuron. Neurons from one layer cannot talk to each other or to the neurons in arbitrary layers; they may only send their signal to the neurons in the succeeding layer. A neuron receives the weighted sum of the activation of all the neurons in the previous layer, adds an internal parameter known as the bias and maps this sum to a value computed by an activation function (e.g. sigmoid, hyperbolic tangent, rectilinear, softmax). This model can be stated mathematically by the following equation:
\begin{equation}\label{eq:activation}
a_{i}^{l} = f(\sum_j{a_{j}^{l-1} w_{j \rightarrow i}^{l} }  + b_{i}^{l} ).
\end{equation}
Here, $a_{i}^{l}$ is the activation of the neuron in hand (i.e. the $i$'th neuron in the $l$'th layer), $f$ is the activation function of this neuron, $a^{l-1}_j$ is the activation of the neuron $j$ in layer $l-1$ (the previous layer), $w_{j \rightarrow i}^{l}$ is the synaptic weight connecting $i$'th neuron in layer $l$ to the $j$'th neuron in layer $l-1$, and $b_{i}^{l}$ is the bias of the neuron to adjust its activation sensitivity. The first layer, i.e. the input layer, in a deep learning neural net acts as a sensory layer, analogous to the retina. As it gets analyzed, the information from the input layer travels through multiple layers until it reaches the final layer called the classification layer. Each class of images corresponds to a classifying neuron. In our case, we have a neuron corresponding to unlensed and another to lensed images. The neuron with the highest output determines which class an input image is placed in.

A neural net learns how to classify images by adjusting the weights between its neurons and the biases within them, having one goal in mind: minimizing the loss function $C(\boldsymbol{x}, \boldsymbol{y})$. The loss function, sometimes called the cost function, can take many forms but it has to captures the misfiring of the classification neurons, i.e. the deviation between the target class versus the predicted class. This is why such algorithms are known as supervised learning algorithms, in contrast to unsupervised techniques. A common choice for the loss function is the cross-entropy loss function with the following form \citep{nielsen}:
\begin{equation}\label{eq:cost1}
\resizebox{.85\hsize}{!}{$ C(\boldsymbol{x}, \boldsymbol{y}) =\sum_{j =\text{unlensed}, \quad \\  \text{lensed}}{y_j \ln a_{j}^{L} + (1-y_j) \ln (1 - a_{j}^{L})} $}.
\end{equation}
$a_{j}^{L}$ is the activation of neurons in the final (classifying) layer. $\boldsymbol{x}$ is the input data in the vector form and  $\boldsymbol{y}$ represents the desired activations of the two classifying neurons. Of course, this function depends on the architecture of the neural net, weights, and biases, but they have not been expressed explicitly. As an example, if an image is a lens, its target output has to be $(0.0, 1.0)$, meaning the activation of the unlensed neuron should be zero and the activation of the lensed neuron should be unity. During the training process of a neural net, images from a training dataset are presented to the network and the weights and the biases are adjusted to minimize the loss function for those images. The parameter space is massive and a change in one of the parameters of a neuron will affect the activation of a series of neurons in other layers. The first challenge is solved by minimizing algorithms such as the stochastic gradient descent (SGD) and the second one is solved via back-propagation. We won't go in the details of these two techniques, but it worths emphasizing on the stochasticity of SGD. Stochasticity refers to randomly selecting images from the training set and bundling them in one batch. The loss function for the batch is the average of the loss function for individual images. It is very important to use a batch rather than training the neural net with one image at a time. Loosely speaking, using a batch would drastically improve classification accuracy of the network since neurons learn the features in images rather than memorizing examples.

ConvNets are a class of neural networks with multiple convolutional layers. A convolutional layer consists of a set of convolving neurons (on the order of 10 neurons) which can be connected to a small rectangular region of an image.  The set of weights of a convolving neuron is known as a filter and are subject to change as the ConvNet learns. A filter scans an entire image by striding (convolving with specified steps) over the image and assembling its output into an image knows as a feature map. Feature maps contain information such as texture and edges. See Figure \ref{fig:filters1} as an example of a set of filters in a LensFlow convolutional layer. Two extracted feature maps of a physical lens (not simulated) have been shown in Figure \ref{fig:feature_maps}. 

\begin{figure} 
    \includegraphics[width=0.5\textwidth]{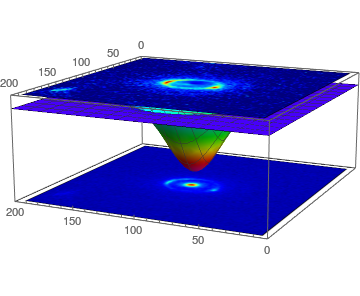}
    \centering
  \caption{Anti-Gaussian image enhancement technique. The bright source at the center of the original (bottom) image has been deemed by an anti-Gaussian function (middle) which attenuates central pixel values (see Equation \ref{eqn:antigaussian}). In addition,  gamma correction has been employed to adjust the contrast, resulting in an enhanced (top) image where the arcs stand out. Such enhanced and normalized images are the inputs of the ConvNet.}
  \label{fig:antigaussian}
\end{figure}

\section{Methodology}\label{methodology}
\subsection{Training and Testing Datasets}\label{datasets}
Neural networks learn to classify data by learning from examples, referred to as the training dataset. A training dataset usually contains a few thousands of pre-classified images. Our training dataset contains $15,000$ unlensed sources and $15,000$ lensed sources. To make this dataset, we used {\sc SExtractor} to obtain a catalog of sources from a few high-quality central tiles as well as a few low-quality edge tiles to include noisy images and artifacts. A cutout of $200 \times 200$ pixels was made around the identified sources and stored individually, labeled as unlensed. After training and finalizing the architecture of the ConvNet, using a source extraction software is not necessary since an entire tile can be divided into smaller tiles and scanned to identify locations with high lensing probability.

Creating lenses are more challenging and more involved. To create these lenses, we had two options; we could either artificially boost up the number of know lenses or simulate them. We tried both methods. For the first method, we used 17 out of 18 lenses discussed in \citet{cosmos2011} paper.\footnote{One of the lens candidates from \citet{cosmos2011} is more similar to a tidal interaction.}
Since the number of known lenses is very limited, we formed $200 \times 200$-pixel cutouts by applying all combination of the following transformations: (a) Rotating from 0 to 360 degrees in steps of 30 degrees, (b) Shifting the center by $0$, $\pm10$, and $\pm 25$ pixels in horizontal and vertical directions, (c) Scaling by 0.8, 1, 1.1 and (d) Adding a small Gaussian noise after transformation. This combination results in 900 transformed images per lens, boosting up our training dataset to fifteen thousand. 

\begin{figure*}\label{fig:convnet}
\includegraphics[scale=0.35]{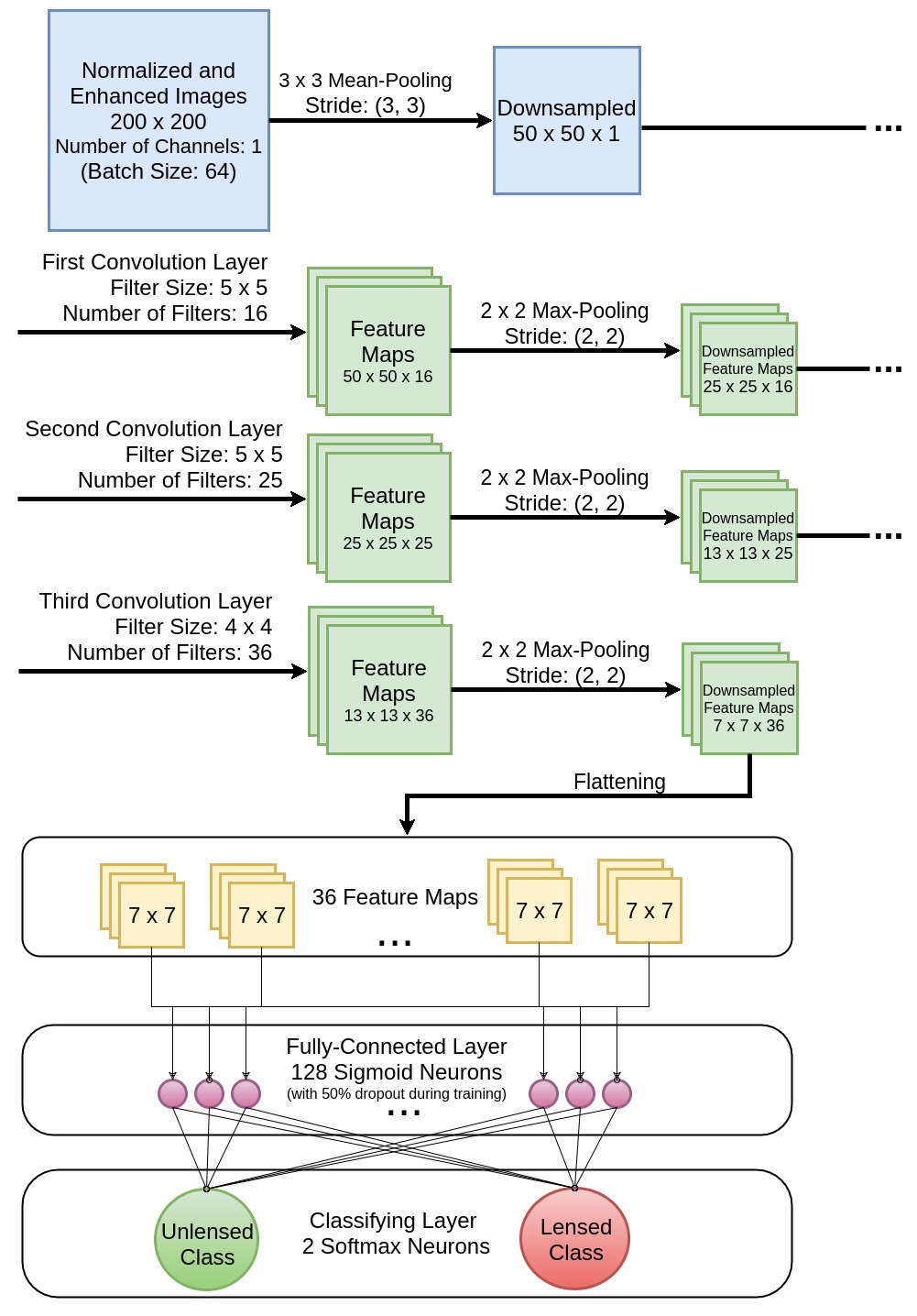}
\centering
\caption{Representation of data flow through the ConvNet layers. The data is down-sampled and fed to three convolutional-max-pooling layers. The data is then flattened into an array and is fed to 128 sigmoid neurons. These neurons are then connected to 2 classifying softmax neurons. These last two layers are responsible for learning the difference between unlensed and lensed images buy comparing and analyzing the feature maps. }
\label{fig:convnet}
\end{figure*}

\begin{figure*}
\begin{tabular}{ll}
\includegraphics[width=0.5\textwidth]{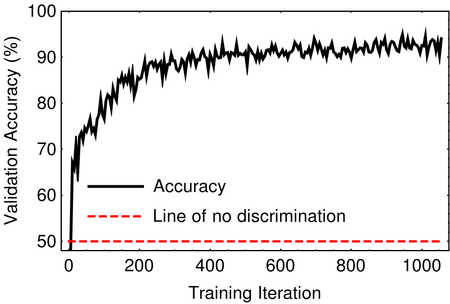}\label{fig:acc}
&
\includegraphics[width=0.5\textwidth]{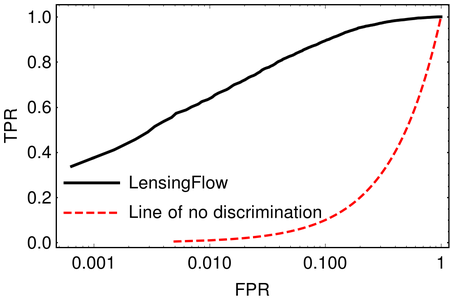}\label{fig:roc}
\end{tabular}

\caption{Left: Validation accuracy vs number of training iterations. Before learning, the ConvNet randomly classifies the images resulting in a $50\%$ accuracy (since there are an equal number of unlensed and lensed images in the testing dataset). After approximately $500$ iterations, the accuracy asymptotes to about $ 91\%$. Higher accuracies could be achieved with larger training datasets and more convolutional layers which are limited by human and by computational resources respectively.
Right: Receiver operating characteristic (ROC) diagram. The black curve shows the trend of the true positive rate verses the false positive rate of LensFlow in the log-linear scale while the red dashed curve shows an untrained classifier.}
\label{fig:acc}
\label{fig:roc}
\end{figure*}

\begin{figure}
    \includegraphics[width=0.45\textwidth]{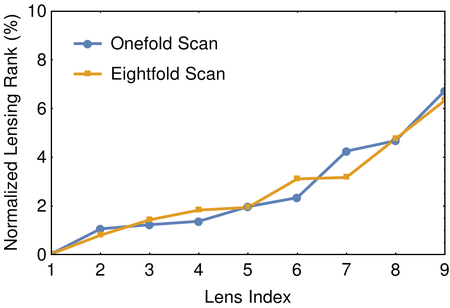}
    \centering
  \caption{Normalized ranking of tracer lenses by LensFlow. LensFlow ranks images based on their lensing probabilities, the highest probability has a rank of 1, the second highest probability has a rank of two, etc. The normalized rank (i.e. absolute rank divided by the total number of unlensed and lensed samples) of top 10 (out of 15) identified COSMOS lenses that were added to 3488 COSMOS sources has been plotted. The blue lines show the result of scanning images once and the orange lines represent scanning the 8 symmetry group transformations of the square applied to each image to artificially increase the number of data points. No significant improvement is detected. This plot indicate LensFlow can place the majority of the physical lenses in the top $8\%$ minority.}
\label{fig:top_ranks}
\end{figure}

However, the training dataset using the transformation method does not contain a wide range of lenses and lenses with different structures might be missed. For this reason, we have also created a simulated training dataset using {\sc LensTool} \citep{lenstool1, lenstool2, lenstool3}. {\sc LensTool} receives input parameters of the lensing model and generates a lensed image of the background galaxy without a foreground lensing source. We will refer to these as arcs. Lens model parameters such as redshift of the background and foreground galaxies, their ellipticity, orientation, and their relative positions were randomly changed to create a wider range of arcs, from complete Einstein rings with different radii to arcs with different orientations and sizes. The code for generating these arcs which includes the range of model parameters will be included in our online distribution. After generating a wide range of arcs, they were merged with raw images. Both images, the raw sources, and the simulated arcs, were normalized and the pixels of the arc images were multiplied by a random number between $0.3$ and $0.8$ so a range of relative arc to foreground luminosity would be captured in the training set. For a limited number of generated lenses, other ranges were used to create extremely faint and extremely bright lenses. $85\%$ of the lenses were generated by selecting elliptical sources while the remaining were sources randomly selected from tile 55. A bias like this is necessary since the foreground of the know COSMOS lenses are all elliptical. After examining the simulated lenses by eye and eliminating lenses with unnatural geometry, an eightfold transformation was performed to increase the number of lenses up to fifteen thousand. These transformations come from 8 elements of the symmetry group of the square, namely: $0\,^\circ$, $90^\circ$, $180^\circ$ and $270^\circ$ rotations, and horizontal, vertical, diagonal and anti-diagonal mirroring. 

\begin{figure} 
    \includegraphics[width=0.5\textwidth]{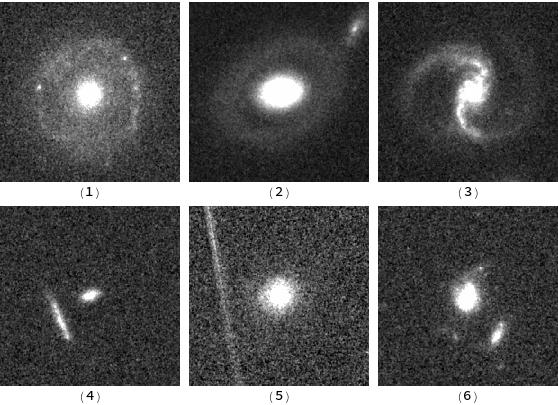}
    \centering
  \caption{Examples of common lens contaminants. Due to their similar visual geometries with lenses, ring galaxies (1-2) and spiral galaxies (3) are often false positive classifications of LensFlow. Galaxies with satellite or tidal interactions (4 and 6) and artifacts from bright sources (5) also contaminate the lens candidates.}
\label{fig:contaminants}
\end{figure}

At the end, 500 lenses and 500 raw images were removed from the training set and placed in a separate set called the testing dataset to measure the performance of the ConvNet as the network gets trained on the training dataset.

\begin{figure*}
\begin{center}
\includegraphics[scale=0.4]{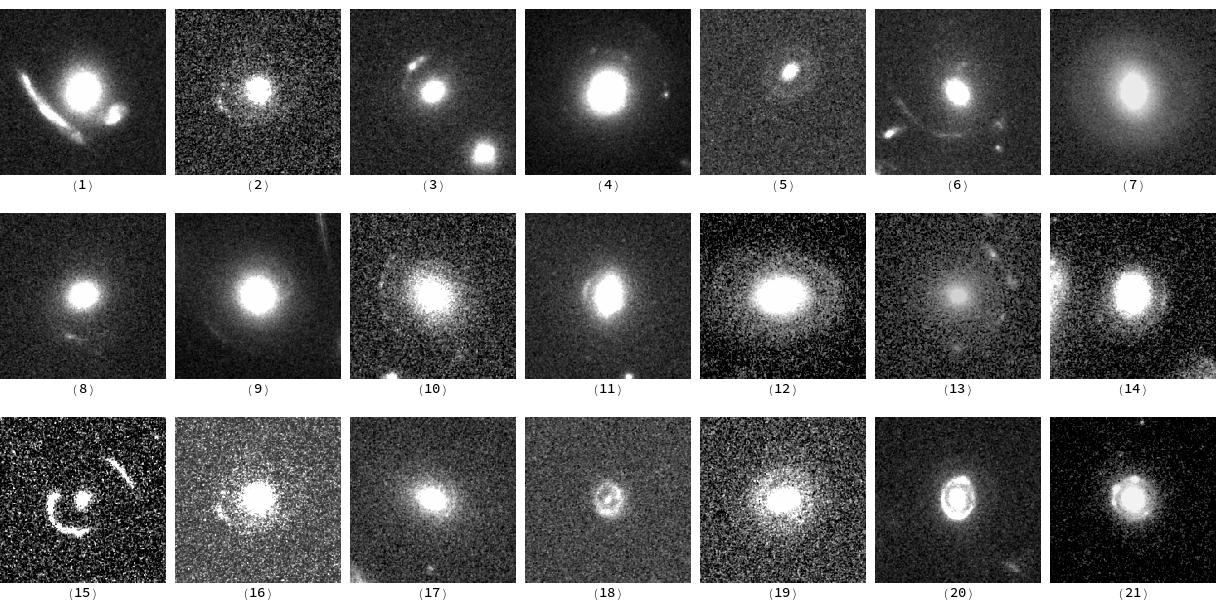}
  \centering
  \caption{Identified COSMSOS lens candidates by LensFlow. These candidates were visually selected from images with the highest lens probability assigned by LensFlow. Candidates (18) and (19) were identified by retraining LensFlow on perfect Einstein rings only. RA and Dec of these lens candidates are listed in Table \ref{tab:lens_coords}.}
  \label{fig:cosmos_lenses}
\end{center}
\end{figure*}

\subsection{Image Normalization and Enhancement}\label{normalization}
Before inputting the images to our neural net, we normalize and enhance them. For deep learning purposes, there are different methods of image normalization to choose from, but not normalizing is not a choice. This is due to the fact that raw images come in a wide range of values. However, a limited number of neurons are not capable of handling such variations. On top of that, the goal of 
a ConvNet is to learn geometric features rather than the statistical properties of the pixels, which could be unwillingly introduced while constructing the training dataset. To address these two issues, all input images must be normalized.  Our method of normalization consists of two steps. In the first step, pixel values are shifted by their average so their new mean would be zero (Equation \ref{eqn:norm1}). In the second step, pixel values in an image are divided by their standard deviation (Equation \ref{eqn:norm2}). 
\begin{equation}\label{eqn:norm1}
{p_{ij}} \leftarrow {p_{ij}} - \frac{\sum_{i  j}{p_{ij}} }{i_{max}  j_{max}}
\end{equation}
\begin{equation}\label{eqn:norm2}
{p_{ij}} \leftarrow  \frac{p_{ij}}{\sqrt{ \sum_{i  j}{p_{ij}^2}}}
\end{equation}

$p_{ij}$ stands for the value of the pixel at row $i$ and column $j$. $i_{max}$ and $j_{max}$ indicate the total number of rows and columns, respectively. In order to improve our classification accuracy, we made the arcs stand out by apply an anti-Gaussian function (Equation \ref{eqn:antigaussian}) on each image. This function will attenuate the central pixels which belong to the foreground galaxy (see Figure \ref{fig:antigaussian}).  The width of this function is about a quarter of the cutout width.

\begin{equation}\label{eqn:antigaussian}
{p_{ij}} \leftarrow (1-e^{-((i-100)^2 +(j-100)^2)/1000}) {p_{ij}}
\end{equation}

At this step, the images are ready for contrast adjustments which is done using the gamma correction. The pixels must be shifted up in order to eliminate zero or negative pixel values.

\begin{equation}\label{eqn:gamma}
{p_{ij}} \leftarrow (p_{ij} + 1.01 \min(\boldsymbol{p}) )^{0.15}
\end{equation}
The normalization steps are applied again and at the end, the negative pixels are dropped.
\begin{equation}\label{eq:gamma}
{p_{ij}} \leftarrow \min(p_{ij}, 0)
\end{equation}
This improves the image quality by removing low-intensity pixels which mostly include noise. Now images may be inputted to the neural net for training or for classification of new data. See Figure \ref{fig:antigaussian} for an example of image normalization and enhancement. 

\begin{table*}
\begin{center}
\caption{Catalog of displayed lenses in Figure \ref{fig:cosmos_lenses}.}
\begin{tabular}{ l  c  c  c  c  c  }
\hline
Lens & Object ID & Right Ascension & Declination & Einstein  Radius  &  Magnitude$^{\dagger}$ \\ 
 & & (deg) & (deg) & (arcsec) & (AB) \\
\hline
$1^*$ & COSMOS0108+4029 & 150.2849 & 2.6749 & 1.51 & 19.08 \\  
$2  $ & COSMOS5844+3753 & 149.6860 & 1.6315 & 1.53 & 21.65 \\ 
$3^*$ & COSMOS5831+4334 & 149.6298 & 1.7263 & 1.17 & 20.76 \\  
$4  $ & COSMOS0142+5447 & 150.4285 & 1.9133 & 1.49 & 18.67 \\  
$5^*$ & COSMOS0124+5121 & 150.3522 & 1.8559 & 0.79 & 22.31 \\  
$6^*$ & COSMOS0047+5023 & 150.1986 & 1.8398 & 1.71 & 20.64 \\  
$7  $ & COSMOS0027+0513 & 150.1140 & 2.0871 & 1.80 & 18.87 \\  
$8^*$ & COSMOS0208+1422 & 150.5355 & 2.2396 & 1.54 & 20.13 \\  
$9^*$ & COSMOS0056+1226 & 150.2366 & 2.2072 & 2.04 & 18.78 \\  
$10  $ & COSMOS0237+2652 & 150.6560 & 2.4478 & 1.88 & 20.81 \\  
$11^*$ & COSMOS0012+2015 & 150.0526 & 2.3377 & 0.92 & 19.37 \\  
$12  $ & COSMOS5918+1911 & 149.8272 & 2.3198 & 1.89 & 19.86 \\  
$13  $ & COSMOS0211+2955 & 150.5486 & 2.4986 & 1.91 & 20.82 \\  
$14  $ & COSMOS0157+3510 & 150.4879 & 2.5863 & 1.20 & 20.68 \\  
$15^*$ & COSMOS0018+3845 & 150.0770 & 2.6460 & 1.26 & 23.62 \\  
$16  $ & COSMOS5844+3753 & 149.6860 & 1.6315 & 1.45 & 21.65 \\  
$17  $ & COSMOS0247+5601 & 150.6974 & 1.9337 & 0.82 & 20.24 \\  
$18  $ & COSMOS0229+1441 & 150.6260 & 2.2448 & 0.51 & 26.67 \\  
$19  $ & COSMOS5954+1128 & 149.9791 & 2.1913 & 0.97 & 21.19 \\  
$20^*$ & COSMOS0038+4133 & 150.1595 & 2.6927 & 0.55 & 20.51 \\  
$21^*$ & COSMOS5921+0638 & 149.8407 & 2.1106 & 0.72 & 20.43 \\ 
\hline
\label{tab:lens_coords}
\end{tabular}
\end{center}
\centering
\noindent{\footnotesize Notes. ${}^*$: also identified in \citet{cosmos_paper}. $^{\dagger}$: from ACS i-band.}
\end{table*}

\subsection{Architecture of LensFlow ConvNet}\label{convnet arch}

The architecture of the data determines the dimensionality of the ConvNet layers. We use $200 \times 200 \times 1$ images where $1$ indicate the number of color channels\footnote{In this paper and in our code, we have adapted the $N \times H \times W \times C$ format from {\sc TensorFlow} where $N$, $H$, $W$, and $C$  stand for number of input images in a batch, height, width, and number of color (or feature) channels.}. In our code, we have localized these parameters as input variables to ease the transition from one to multiple color bands. Classifying lenses with multiple color bands will be easier and more accurate since foreground and background sources have a color contrast. However, we have chosen to use one color channel so our algorithm can be sensitive to geometry rather than color contrast in order to expand its applicability to a wider range of bands as well as eliminating its need for multi-band images when unavailable. As it can be seen in Figure \ref{fig:convnet}, after normalizing and enhancing these single-channeled images, LensFlow applies an average-pooling of kernel $4 \times 4$ and stride of $4$. This means that the image is divided into $4 \times 4$ segments and the average of each segment will become a pixel of the down-sampled output image. These down-sampled images are then fed to the first convolutional layer which consists of sixteen $5 \times 5$ filters. The hyperbolic tangent function is chosen as the activation function for the neurons in this layer as well as the upcoming convolutional layers. This layer outputs sixteen feature maps which should be interpreted as one image with 16 feature channels, i.e. a $50 \times 50 \times 16$ image using our notation above. These feature maps are down-sampled using a max-pooling layer of size $2 \times 2$ and stride of $2$ resulting in a $25 \times 25 \times 16$ reduced feature map. This means the input of the max-pooling layer is divided into $2 \times 2$  tiles whose maximum values are selected to form down-sampled feature maps. This $25 \times 25 \times 16$ feature map is inputted to the second convolutional layer with twenty-five $5 \times 5$ filters where each filter will scan all channels by combining them linearly, hence outputting a $25 \times 25 \times 25$ feature map to be down-sampled to $13 \times 13 \times 25$. Another convolutional layer with thirty-six $4 \times 4$ filters paired with a max-pooling layer is followed, reducing the data size to a $7 \times 7 \times 36$ feature map. This information is flattened into a 1-D array where it is fully connected to 128 sigmoid neurons. The output of these neurons is inputted to the classifying layer with 2 neurons with the softmax activation function:
% xxx: source for combining them linearly
\begin{equation}
 \sigma_c(\mathbf{Z}) = \frac{e^{Z_c}}{e^{Z_{\text{unlensed}}} + e^{Z_{\text{lensed}}}} \, .
\end{equation}
Here, each component of $\mathbf{Z}$ is the total weighted input plus the bias of each of the classifying neurons, $z_c = \sum_j{a_{j}^{L-1} w_{j \rightarrow c}^{L} }  + b_{c}^{L} $ . $c$ specifies whether we are talking about the unlensed neuron or the lensed neuron. This function enhances the training process and it will assign a pseudo-probabilistic number to each class (since the probabilities across all classes add up to 1). The higher $\sigma_{c = \text{lensed}}$ for an image, the higher its probability to be a lens which can be used to rank the images based on their lensing probability.

\subsection{Measuring Performance}\label{acc}

To optimize our ConvNet, we have chosen a cross-entropy function as our loss function which we minimize using Adam's Optimizer. During the training phase, 64 unlensed and 64 lensed images were placed in a batch of 128 images. This combination technique will prevent under- or over-representation of classes even if the training size for different classes contain a different number of examples. This process was iterated one thousand times. An example of accuracy as a function of the number of iterations has been graphed in Figure \ref{fig:acc} where the accuracy plateaus at $91.5\%$. To obtain this accuracy, any image with a lensing probability lower/higher than $0.5$ was labeled lensed/unlensed. These predicted labels were compared against the true label of each image. Thus, accuracy is defined as the number of correct classifications over the number of images in the testing set. 
\begin{equation}\label{eq:acc}
\text{accuracy} \equiv \frac{N(\text{lensed}|p_\text{lens} > 0.5)}{ N(\text{lensed}) + N(\text{unlensed})}
\end{equation}
This result was achieved with a "ReLu"\footnote{A rectified linear unit is a neuron with the following common activation function:$ f(x) = \max(0, x)$.} neurons in the convolutional layers. After switching to hyperbolic tangent, our results improved to $96\%$ meaning that for every 100 test images, 4 were misclassified. If the testing dataset contains simulated lenses, the reader should not translate these accuracies as a measure of physical lens classification performance, neither for our paper nor for others. Unlike physical lenses, simulated lenses may contain statistical artifacts and are limited in shape. Hence, the true performance of a classifier should be measured with its ability to locate physical lenses in a sufficiently large field.

To optimize computational time while addressing this issue, we have mixed 15 known lenses from the COSMOS field \citep{cosmos2011} with 3.5 thousand images from the same field and have assigned a lensing probability to each image that was obtained from the output of the lens class neuron. These images were ranked based on their lensing probabilities. The relative ranking for these lenses, i.e. their rank divided by the total number of scanned images, has been plotted in Figure \ref{fig:top_ranks}. We have also tried eightfold scanning where the probabilities of eight transformations discussed above (Section \ref{datasets}) have been summed to improve the accuracy. This technique does not show an improved ranking of the lenses. The majority of the lenses fall under top $8\%$ where they can be further examined by eye. In the next section, we will discuss the images that contaminate the high-rank candidates. We will also discuss the remedies to further separate lenses from other sources.

\section{Results \& Discussion}\label{cosmos}
A catalog of the strong gravitational lenses in the COSMOS field has been generated previously \citep{cosmos_paper} by looking at early type bright galaxies in the redshift range of $0.2 \leq z \leq 1.0$ and visually inspecting and cataloging sixty high and low-quality lens candidates. In contrast, we have examined all sources in {\it HST}/ACS i-band of the COSMOS field that are more extended than 200 pixels and are $1.5 \sigma$ brighter than the background, i.e. 230,000 images. After scanning these images with LensFlow, we inspected the top outputs and selected 21 as good quality lens candidates which are presented in Figure \ref{fig:cosmos_lenses}. Their coordinates and other physical parameters are also listed in Table \ref{tab:lens_coords}. The lens candidates previously identified by \citet{cosmos_paper} are marked in Table 1. Among sixty lens candidates presented in \citet{cosmos_paper}, 25 were more extended than our image cutout size of $200 \times 200$ pixels which translates to larger than $3 \arcsec \times 3 \arcsec$ (e.g. COSMOS0208+1422, COSMOS0009+2455) while some others were considered as lens contaminates or low quality lenses rather than as good lens candidates during our visual selection process (e.g. COSMOS0055+3821). To resolve the former issue, a secondary scan can be performed by down-sampling the field to half its size to include larger lenses. 

The main contaminants of our high lensing probability images fall into the main categories of, spiral galaxies with contamination arising from spiral arms and/or ring-like structures, tidally interacting galaxies and satellite/nearby galaxies in a lens-like configuration and image artifacts. Examples of these contaminants are shown in Figure \ref{fig:contaminants}. Given that these contaminants are similar to many of our training examples and that it is also time-consuming to separate them by human eye we created another training dataset which would eliminate such contaminants by only looking for perfect Einstein rings. After another scan of the field, we were able to identify the Einstein ring shown in the last two panels of Figure \ref{fig:cosmos_lenses}. 

These indicate that perhaps, rather than increasing the number of layers in series, we must aggregate parallel ConvNets, each one trained to extract lenses from one class of contaminants only. Furthermore, reducing the down-sampling factor would, in turn, increase the resolution of the input images which might increase the gap between spirals and lenses at a greater computational cost. And perhaps, it would be more efficient to have separated training datasets for visually distinct lenses, as tried above, rather than having one large training dataset with a variety of lenses. We will study these algorithms in our future publications.

\section{Summary}\label{summary}
In this paper, we have emphasized the importance of gravitational lenses in the field of cosmology and have presented an introduction to neural networks including ConvNets. Furthermore, we have laid out the procedure for constructing simulated images for training and testing LensFlow. The importance and details of our normalization and enhancement methods have been discussed.  Then, we have discussed the architecture of LensFlow, its accuracy on test data, and its performance on real data. The latter was done by scanning \textit{HST}/ACS i-band images of the COSMOS field and listing the lens candidates that were visually separated from images with large assigned lens probability by LensFlow.

\section*{Acknowledgement}\label{acknowledgement}

Financial support for this paper was provided by NSF grant AST-1213319 and GAANN P200A150121. Figure \ref{fig:neuron} and \ref{fig:convnet} were generated using \url{www.draw.io}. The backbone of our algorithm was initially inspired by Hvass Laboratories on Github \citep{HvassLabs}. Figure \ref{fig:filters1} was plotted using his code as well. We would like to thank Pierre Baldi for valuable feedback. We would also like to thank Noah Ghotbi and Aroosa Ansari for their assistance. 

\bibliography{refs.bib}

\clearpage

\appendix
\section{Remarks on Present and Future of LensFlow Code} \label{appendix1}
LensFlow has been written in Python 3.5.2 and to enhance the user interface, it was developed in a Jupyter Notebook environment which enables the user to easily modify the code, plot, and read the documentation alongside the code. LensFlow relies on Keras, a high-level neural networks API, written in Python and capable of running on top of {\sc TensorFlow}\citep{Chollet2015}. Our product consists of three main notebooks: Development notebook, Arc Maker notebook, LensFlow notebook. 

The Development notebook contains the necessary code for file management, to automate source extraction, to merge raw images with simulated arcs, to create training and testing datasets, for automated handling of 81 COSMOS tiles, and includes other helpful functions. Currently, LensFlow is in its development stage and requires a list of $200 \times 200$-pixel FITS cutouts. In future versions, we will eliminate this need and users could provide a large FITS  tile with/without a catalog of identified sources which would require less/more time to scan the data. This can be done by convolving the ConvNet itself over a FITS tile and identify spots with high lensing probabilities. To make this fully functional, the training dataset must include off centered and cropped sources. However, for the purposes of training and optimizing, isolated cutouts are more useful since sources from different tiles can be mixed without facing memory shortage and since sources have to be accessed again and again as one would change the architecture of the ConvNet.

Arc Maker notebook automates the arc creation process and uses {\sc LensTool}. LensFlow notebook contains the ConvNet and can be used to specify the architecture of the ConvNet, to train or to test the ConvNet, and to search through new data for lens candidates. This notebook also contains the functions that we have used to generate some of the plots in this paper as well as the feature to view multiple top lens candidates in one screen using DS9. This feature will help the user to quickly examine the output of the LensFlow. Due to the restrictions that {\sc TensorFlow} and {\sc LensTool} apply, LensFlow would only function in a Linux operating system. 

\end{document}